\documentclass[a4paper,11pt]{article}
\pdfoutput=1 

\usepackage{jcappub} 

\usepackage[T1]{fontenc} 

\title{\boldmath Generalized Gibbons-Werner method for stationary spacetimes}

\author[a,b,c,d]{Yang Huang}
\author[c,d,e,1]{Zhoujian Cao \note{Corresponding author.}}
\author[a,b]{Zhenyan Lu}

\affiliation[a]{School of Physics and Electronic Science, Hunan University of Science and Technology, Xiangtan 411021, China}
\affiliation[b]{Key Laboratory of Intelligent Sensors and Advanced Sensing Materials of Hunan Province, Hunan University of Science and Technology, Xiangtan 411021, China}
\affiliation[c]{Institute for Frontiers in Astronomy and Astrophysics, Beijing Normal University, \\ Beijing 102206, China}
\affiliation[d]{Department of Astronomy, Beijing Normal University, \\ Beijing 100875, China}
\affiliation[e]{School of Fundamental Physics and Mathematical Sciences, Hangzhou Institute for Advanced Study, UCAS,\\ Hangzhou 310024, China}

\emailAdd{yanghuang55@outlook.com}
\emailAdd{zjcao@bnu.edu.cn}

\abstract{The Gibbons-Werner (GW) method is a powerful approach in studying the gravitational deflection of particles moving in curved spacetimes. The application of the Gauss-Bonnet theorem (GBT) to integral regions constructed in a two-dimensional manifold enables the deflection angle to be expressed and calculated from the perspective of geometry. However, different techniques are required for different scenarios in the practical implementation which leads to different GW methods. For the GW method for stationary axially symmetric (SAS) spacetimes, we identify two problems: (a) the integral region is generally infinite, which is ill-defined for some asymptotically nonflat spacetimes whose metric possesses singular behavior, and (b) the intricate double and single integrals bring about complicated calculation, especially for highly accurate results and complex spacetimes. To address these issues, a generalized GW method is proposed in which the infinite region is replaced by a flexible region to avoid the singularity, and a simplified formula involving only a single integral of a simple integrand is derived by discovering a significant relationship between the integrals in conventional methods. Our method provides a comprehensive framework for describing the GW method for various scenarios. Additionally, the generalized GW method and simplified calculation formula are applied to three different kinds of spacetimes—Kerr spacetime, Kerr-like black hole in bumblebee gravity, and rotating solution in conformal Weyl gravity. The first two cases have been previously computed by other researchers, affirming the effectiveness and superiority of our approach. Remarkably, the third case is newly examined, yielding an innovative result for the first time.}

\begin{document}
\maketitle
\flushbottom

\section{Introduction}\label{introduction}
The motion of objects in strong gravitational field is one of the most significant research topics in general relativity (GR). By regarding an object as a test particle whose gravitational field can be ignored, one can analyze its trajectory via the geodesic equation of the strong gravitational field. The deflection of trajectories is an important aspect in the investigation of the motion of test particles.

Generally, the particle can be classified into two categories: massless particles (photons) and massive particles. For massless particles, the gravitational deflection plays a significant role in verifying GR and other gravitational theories \cite{will2014confrontation}. After Eddington's observation confirmed the deflection of light passing by the Sun \cite{dyson1920ix}, gravitational lensing has been deeply and extensively studied as a powerful tool in various fields of astronomy and cosmology \cite{dodelson2017gravitational}. Numerous approaches have been proposed to calculate the deflection angle for photons \cite{weinberg1972principles,vilenkin1984cosmic,vilenkin1985cosmic,gott1985gravitational,bozza2001strong,bozza2002gravitational,bodenner2003deflection,wucknitz2004deflection,rindler2007contribution,sultana2013contribution,bhattacharya2010light,bhattacharya2011vacuole,cattani2013correct,farrugia2016solar,mishra2018trajectories,ishak2008light,sereno2008influence,sereno2009role,dey2008gravitational,bhattacharya2010bending,gallo2011gravitational,gallo2012peculiar,bozza2015alternatives,boero2018gravitational,crisnejo2018expressions,bisnovatyi2017gravitational,guenouche2018deflection,glavan2020einstein,kumar2020gravitational,jin2020strong,heydari2021bending,panah2020charged,jafarzade2021shadow,atamurotov2021charged}. For massive particles, they can also serve as messengers of the universe. Examples include neutrons, neutrinos, cosmic rays from high-energy celestial events ($\pi$-mesons, µons, K-mesons, etc.), theorized weakly interacting massive particles and axions \cite{patla2013flux}, gravitons. The gravitational deflection of these massive particles provides valuable information about the source, the lens, the background of the trajectory and the particles themselves \cite{liu2019constraining,accioly2004photon, bhadra2007testing,tsupko2014unbound,liu2016gravitational,he2016gravitational,he2017analytical,pang2019gravitational,li2019gravitational}. 

The GW method, initially proposed by Gibbons and Werner in 2008 \cite{gibbons2008applications} and subsequently developed by researchers in recent years \cite{werner2012gravitational,ishihara2016gravitational,ono2017gravitomagnetic,crisnejo2018weak,jusufi2018gravitational,li2020thefinitedistance}, contributes to the calculation and understanding of the deflection angle for both massless and massive particles from the geometric perspective. The basic scheme of the GW method involves: (a) constructing a reduced three-dimensional space which can be used to describe the particle's motion in four-dimensional spacetimes, (b) establishing a two-dimensional Riemannian manifold to describe the motion in the equatorial plane based on the reduced three-dimensional space, (c) defining an integral region on the two-dimensional Riemannian manifold (typically enclosed by four curves: the particle's trajectory, an auxiliary circular arc, a radial outward curve passing through the source, and a radial outward curve passing through the observer), and (d) applying the GBT to the integral region to express the deflection angle in terms of geometric quantities. It should be remarked that, while there are differences in the technical details between the calculations for photons and massive particles, the calculation process and result for massive particles can reduce to those for photons when the rest mass approaches zero and the velocity approaches the speed of light. Throughout this paper, the term "particles" refers to both massless and massive particles.

The practical calculation process of the GW method requires different techniques depending on the specific situation. In this paper, the term "GW method" is a general reference to all methods that calculate the deflection angle of particles based on the aforementioned scheme. Although the original GW method proposed by Gibbons and Werner is based on the static spherically symmetric (SSS) spacetime, it has been extended to SAS scenarios with three different techniques \cite{werner2012gravitational,ono2017gravitomagnetic,jusufi2018gravitational}. Among these techniques, the one proposed by Ono, Ishihara, and Asada (referred to as GWOIA method) \cite{ono2017gravitomagnetic} is the most powerful and widely used due to its flexibility and straightforwardness. Specifically, in GWOIA method, the equatorial plane of the Riemannian part of a Rander-Finsler metric is selected as the two-dimensional Riemannian manifold, and the deflection angle is expressed in terms of the integral of the Gaussian curvature and the geodesic curvature. More related works for SAS spacetimes using GWOIA method can be found in \cite{ono2018deflection,ovgun2018light,ovgun2018shadow,haroon2019shadow,ono2019deflection,ono2019effects,kumar2019shadow,li2020equivalence,li2020finite,li2020thefinitedistance,li2021kerr,li2021kerrnewman,li2022deflection,huang2023finite,pantig2023testing}.

However, the existing works studying the deflection angle for SAS spacetimes using the GWOIA method face two main challenges. First, the auxiliary circular arc is located at the infinite region, thus the resulting infinite integral region is ill-defined for certain asymptotically nonflat spacetimes, such as the Kerr-de Sitter spacetime \cite{dewitt1973black} and the rotating solution in conformal Weyl gravity \cite{mannheim1991solutions,varieschi2014kerr}, which encounter singularities as the radial coordinate approaches infinity. Second, the calculation formula contains a double integral and a single integral, and the quantities involved (the Gaussian curvature, the geodesic curvature, and the upper and lower bounds of integrals) are very complex, making the computation cumbersome. 

In this paper, we simultaneously address these two challenges by proposing a generalized GW method and giving its corresponding simplified calculation formula. The work in this paper is based on the discovery of an important relation between the integral of the Gaussian curvature over the integral region and the integral of the geodesic curvature along the auxiliary circular arc. Moreover, an interesting development emerged during the review of this paper—Ishihara et al. noticed our manuscript on arXiv and established the equivalence between our generalized GW method and the GWOIA method \cite{takahashi2023equivalence}. Their discovery significantly enhance our confidence in the robustness of our methodology.

The remainder of this paper is organized as follows. In Sec.~\ref{sec-2}, we provide a review of the GBT, the Jacobi-Maupertuis Randers-Finsler (JMRF) metric, and the finite-distance deflection angle. Sec.~\ref{sec-3} presents a brief introduction to the GWOIA method. In Sec.~\ref{sec-4}, we propose the generalized GW method and derive the corresponding simplified calculation formula. In Sec.~\ref{sec-5}, we demonstrate the validity and superiority of our method and formula by performing calculations for different spacetimes. Finally, we conclude the paper in Sec.~\ref{conclusion}. In the upcoming sections, we adopt the spacetime signature ($-,+,+,+$), and geometric units where the gravitational constant $G$ and the speed of light are set to one, i.e. $G=1$ and $c=1$.

\section{GBT, JMRF metric, and Finite-distance deflection angle}
\label{sec-2}
\subsection{GBT}
\begin{figure}[!ht]
  \centering
	\includegraphics[width=0.6\columnwidth]{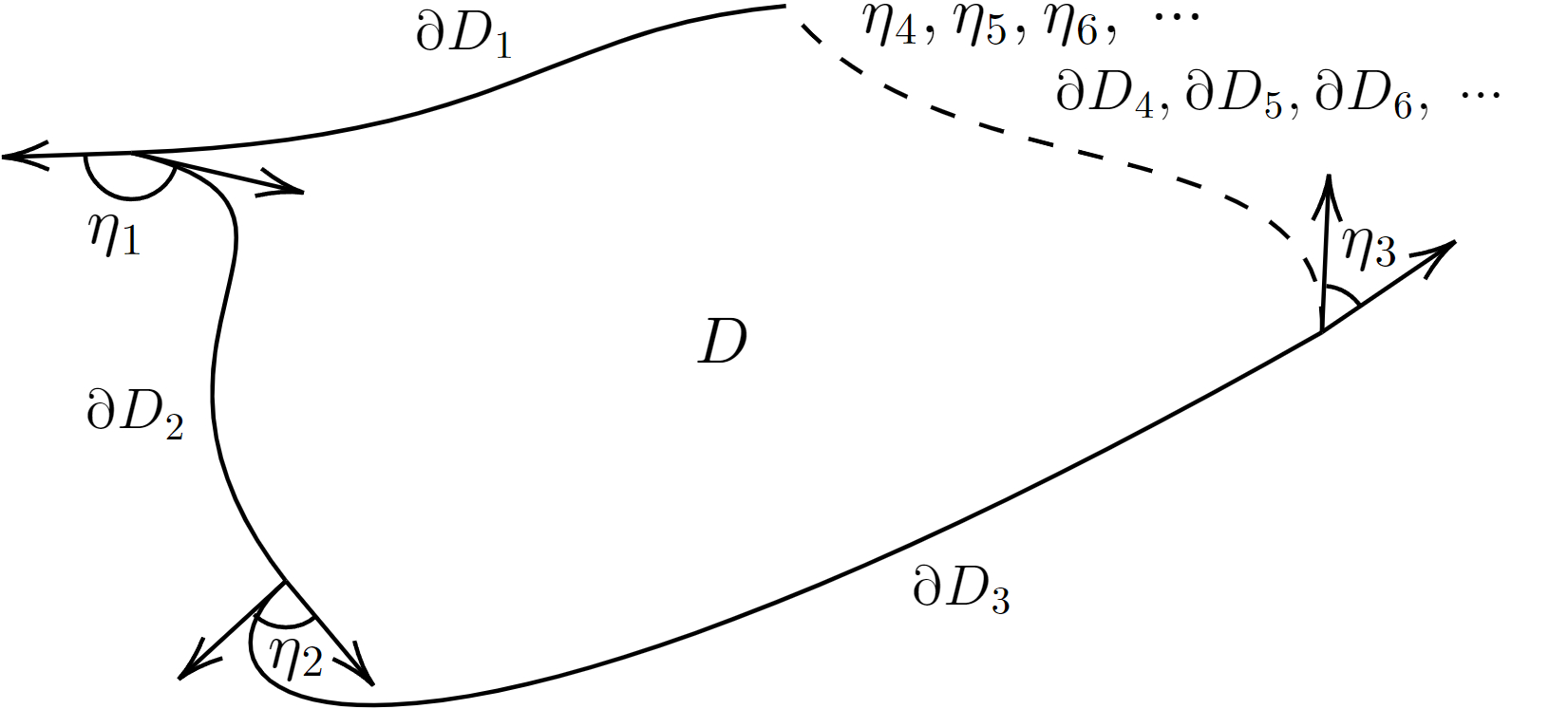}
	\caption{ A region $D$ with boundary $\partial D=\cup_{i} \partial D_{i}$. The jump angles of $D$ are denoted by $\eta_i(i=1,2,\cdots)$ in the positive sense.}
	\label{fig-1}
\end{figure}

Let $D$ be a compact and connected region on a two-dimensional Riemannian manifold (as depicted in Fig.~\ref{fig-1}).
The boundary of $D$, denoted as $\partial D$, consists of piecewise smooth components $\partial D_i$ ($i=1,2,\cdots$), and the jump angles at each vertex are represented by $\eta_i$ in the positive sense. Then the GBT can be expressed as \cite{manfredo1976carmo}
\begin{equation}
    \iint_{D} K \mathrm{d} S+\sum_i\int_{\partial D_i} \kappa \mathrm{d}l+\sum_{i} \eta_{i} = 2 \pi \chi(D),
    \label{gbt}
\end{equation}
where $K$ and $\mathrm{d}S$ are the Gaussian curvature and area element of $D$, respectively; $\kappa$ and $\mathrm{d}l$ represent the geodesic curvature and line element of $\partial D$, respectively; $\chi (D)$ denotes the Euler characteristic number of $D$.

Eq.~\eqref{gbt} establishes a fundamental relation between the integral of curvature quantities and the Euler characteristic of the region $D$. The GBT serves as a powerful tool for analyzing the geometric property of surfaces and their topological characteristics.

\subsection{JMRF metric}
The JMRF metric, constructed by Chanda $et\ al.$ in 2019 \cite{chanda2019jacobi}, is of great significance in geometrodynamics and provides a framework for studying the particle's motion in stationary spacetimes. Consider a coordinate $(t,\boldsymbol{x})$ where $\left(\partial/\partial t\right)^a$ is the Killing vector corresponding to the stationary property of the spacetime, the metric of the stationary spacetime states
\begin{equation}
    \mathrm{d}s^2 = g_{00}(\boldsymbol{x}) \mathrm{d}t^2+ 2 g_{0i}(\boldsymbol{x}) \mathrm{d}t\mathrm{d}x^i + g_{ij} (\boldsymbol{x}) \mathrm{d}x^i \mathrm{d}x^j.
    \label{ds2}
\end{equation}
Then the corresponding JMRF metric reads \cite{chanda2019jacobi}
\begin{equation}
    \mathrm{d}\tilde{s} = \sqrt{\alpha_{ij} \mathrm{d}x^i \mathrm{d}x^j} + \beta_i  \mathrm{d}x^i,
    \label{JMRF}
\end{equation}
where the components of the JMRF metric are given by
\begin{align}
    \alpha_{ij} &=  \frac{E^2 +m^2 g_{00}}{-g_{00}} \left(g_{ij}-\frac{g_{0i}g_{0j}}{g_{00}}\right),  \label{alphaij} \\
    \beta_i & = -E \frac{g_{0i}}{g_{00}} \label{betai}.
\end{align}
Here, $m$ and $E$ stand for the rest mass and relativistic energy of the particle, respectively, and we dropped ’$(\boldsymbol{x})$’ from '$g_{00}(\boldsymbol{x})$', '$g_{0i}(\boldsymbol{x})$', and '$g_{ij}(\boldsymbol{x})$' . The Riemannian metric $\alpha_{ij}$ and one-form $\beta_i$ must satisfy the inequality $ \sqrt{\alpha^{ij}\beta_i\beta_j}<1$. 
We denote the three-dimensional space determined by $\alpha_{ij}$, i.e. the Riemannian part of the JMRF metric, as $M^{\left(\alpha 3\right)}$. For a geodesic in the four-dimensional stationary spacetime Eq.~\eqref{ds2}, its spatial projection $\gamma$ is a geodesic in the three-dimensional JMRF space Eq.~\eqref{JMRF} \cite{chanda2019jacobi}. 
Additionally, the JMRF metric encompasses various special cases, including the Jacobi metric applicable to massive particles in SSS spacetimes \cite{gibbons2015jacobi}, the optical Randers-Finsler metric applicable to photons in SAS spacetimes \cite{werner2012gravitational}, and the optical metric applicable to photons in SSS spacetimes \cite{gibbons2008applications}. Researchers have utilized the JMRF metric, as well as its specific forms, to investigate the gravitational deflection of particles \cite{gibbons2008applications,werner2012gravitational,jusufi2016gravitational,jusufi2016light,ishihara2016gravitational,jusufi2017quantum,gibbons2015jacobi,jusufi2017deflection,ishihara2017finite,ono2017gravitomagnetic,jusufi2018deflection,sakalli2018analytical,jusufi2019light,haroon2019shadow,crisnejo2018weak,jusufi2018gravitational,jusufi2019distinguishing,crisnejo2019higher,li2020finite,li2020gravitational,li2020thefinitedistance,li2020circular,huang2022generalized,huang2023extending}, the Kepler orbit \cite{chanda2017jacobi}, the motion of charged particles \cite{das2017motion}, and Hawking radiation \cite{sakalli2017hawking, bera2020hawking}. A comprehensive discussion on the JMRF metric can be found in \cite{chanda2019fermat}. 

For SAS spacetimes, the metric in the Boyer-Lindquist coordinates can be expressed as
\begin{equation}
    \mathrm{d} s^2 = g_{tt}\left(r,\theta \right) \mathrm{d} t^2 + g_{rr}\left(r,\theta \right)\mathrm{d}r^2 + g_{\theta\theta}\left(r,\theta \right)\mathrm{d}\theta^2  + g_{\phi\phi} \left(r,\theta \right) \mathrm{d}\phi^2 + 2 g_{t\phi}\left(r,\theta \right)\mathrm{d}t\mathrm{d}\phi,
    \label{SASmetric}
\end{equation}
and the metric of the corresponding $M^{\left(\alpha 3\right)}$ can be written as
\begin{equation}
    \mathrm{d}{\hat{l}}^2 = \alpha_{rr}\left(r,\theta\right) \mathrm{d}r^2 + \alpha_{\theta\theta}\left(r,\theta\right) \mathrm{d}\theta^2+ \alpha_{\phi\phi}\left(r,\theta\right) \mathrm{d}\phi^2.
    \label{metricMalpha3}
\end{equation}
We focus on the motion in the equatorial plane ($\theta=\pi/2$), and $M^{\left(\alpha 3\right)}$ reduces to a two-dimensional Riemannian space (denoted as $M^{\left(\alpha 2\right)}$ for simplicity). The metric of $M^{\left(\alpha 2\right)}$ reads
\begin{equation}
\mathrm{d}l^2 = \alpha_{rr}\left(r\right) \mathrm{d}r^2 + \alpha_{\phi\phi}\left(r\right) \mathrm{d}\phi^2,
\label{metricMalpha2}
\end{equation}
in which
\begin{equation}
    \alpha_{rr}\left(r\right) =  \frac{E^2+m^2 g_{tt}}{-g_{tt}}g_{rr},  \qquad
    \alpha_{\phi\phi}\left(r\right) =  \frac{E^2+m^2 g_{tt}}{-g_{tt}}g_{\phi\phi}.
    \label{alpharralphaphiphi}
\end{equation}
The corresponding $\beta_i$ states
\begin{equation}
    \beta_\phi\left(r\right) = -E \frac{g_{t\phi}}{g_{tt}}.
    \label{betaMalpha2}
\end{equation}
Here, we dropped '$\left(r,\theta=\pi/2\right)$' from '$g_{tt}\left(r,\theta=\pi/2\right)$', '$g_{rr}\left(r,\theta=\pi/2\right)$', '$g_{\phi\phi}\left(r,\theta=\pi/2\right)$', and '$g_{t\phi}\left(r,\theta=\pi/2\right)$' for simplicity.

Furthermore, the equation of motion for unbound particles moving in the equatorial plane of the SAS spacetime equipped with metric~\eqref{SASmetric} can be written as \cite{huang2023finite}
\begin{equation}
        \left(\frac{\mathrm{d}u}{\mathrm{d}\phi} \right)^2 =  \frac{u^4 \left(g_{t\phi}^2-g_{tt} g_{\phi \phi}\right) \left[  g_{tt} b^2 v^2+2  g_{t\phi}bv + g_{\phi \phi } \left(1+g_{tt}-g_{tt} v^2\right)+g_{t\phi}^2 \left(v^2-1\right)\right] }{g_{rr} ( g_{tt}b v +g_{t\phi})^2}   ,
\label{dudphi}
\end{equation}
where $u= 1/r$, $b$ symbolizes the impact parameter, $v$ is the velocity of particles. 

\subsection{Finite-distance deflection angle}
In 2017, by taking into account the finite distance from a lens object to a light source and a observer, Ishihara $et\ al.$ investigate the finite-distance corrections for the deflection angle of photons \cite{ishihara2016gravitational}. With the GW method, they defined a finite-distance deflection angle and prove it is geometric invariant, namely well-defined.

\begin{figure}[!ht]
    \centering
        \includegraphics[width=0.6\columnwidth]{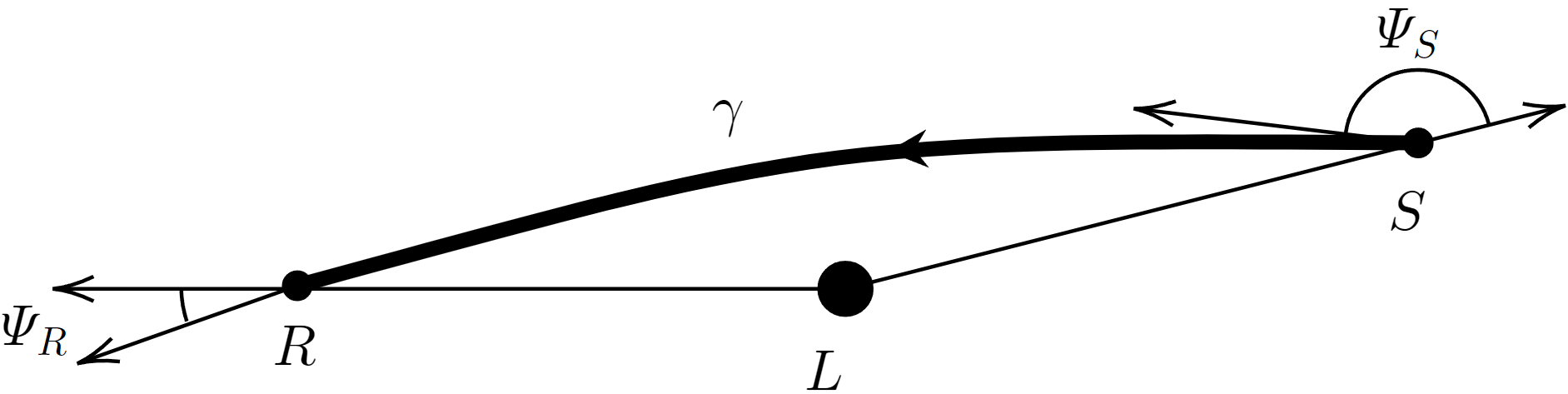} 
        \caption{In a two-dimensional Riemannian manifold, $L$ represents the lens; $S$ and $R$ are the source and the observer of particles, respectively; $\gamma=\overset{\curvearrowright}{SR}$ is the trajectory from $S$ to $R$; $\Psi_S$ and $\Psi_R$ are the angle between the tangent vector along $\gamma$ and the radial outward vector at $S$ and $R$, respectively.}
    \label{fig-2}
\end{figure}

As shown in Fig.~\ref{fig-2},
in a two-dimensional Riemannian manifold, $L$ represents the lens, $S$ and $R$ are the source and the receiver of particles, respectively, $\gamma=\overset{\curvearrowright}{SR}$ is the trajectory from $S$ to $R$, then the deflection angle is defined by \cite{ishihara2016gravitational}
\begin{equation}
    \delta = \Psi_R - \Psi_S + \phi_{RS},
    \label{defAngle}
\end{equation}
in which $\phi_{RS}=\phi_R-\phi_S$ is the increment of the azimuthal coordinate, $\Psi_S$ and $\Psi_R$ are the angle between the tangent vector along $\gamma$ and the radial outward vector at $S$ and $R$, respectively. In the limit as $S$ and $R$ tend to infinity, $\Psi_S=\pi$ and $\Psi_R=0$, the above formula reduces to the usual infinite-distance deflection angle $\delta=\phi_{RS}-\pi$. The definition Eq.~\eqref{defAngle} has been widely employed in the investigation of the deflection of particles \cite{ishihara2017finite,ono2017gravitomagnetic,ono2018deflection,ono2019deflection,haroon2019shadow,kumar2019shadow,crisnejo2019finite,ono2019effects,li2020finite,li2020thefinitedistance,takizawa2020gravitational,li2020circular,li2021kerr,li2021deflection,belhaj2022light,li2021kerrnewman,li2022deflection,pantig2022testing}.

\section{GWOIA method}
\label{sec-3}
In 2017, Ono, Ishihara, and Asada extended the GW method for calculating the finite-distance deflection angle from SSS spacetimes to SAS spacetimes \cite{ono2017gravitomagnetic}. Considering the motion of particles moving in the equatorial plane of an SAS spacetime equipped with the metric~\eqref{SASmetric}, the GWOIA method directly selects the corresponding $M^{\left(\alpha 2\right)}$ as the two-dimensional Riemannian manifold. It is more flexible (can be applied to various scenarios) and more straightforward (easy to understand) compared the osculating Riemannian method \cite{werner2012gravitational} and refractive index method \cite{jusufi2018gravitational}.

Denoting the spatial projection of geodesics in the equatorial plane of the spacetime equipped with metric~\eqref{SASmetric} as $\gamma$, Ono $et\ al.$ demonstrated that $\gamma$ is not a geodesic in $M^{\left(\alpha 2\right)}$. However the deviation of $\gamma$ from the geodesic of $M^{\left(\alpha 2\right)}$ can be described by the corresponding one-form $\beta_i$ \cite{li2020thefinitedistance}. The geodesic curvature of $\gamma$ in $M^{\left(\alpha 2\right)}$ can be evaluated using the following expression \cite{ono2017gravitomagnetic},
\begin{equation}
    \kappa= -\frac{\beta_{\phi,r}}{\sqrt{\hat{\alpha} \cdot \alpha^{\theta\theta}}} ,
    \label{cediqulvcal}
\end{equation}
where $\beta_\phi$ can be derived by Eq.~\eqref{betaMalpha2}, the comma denotes the derivative, $\hat{\alpha}$ is the determinant of metric~\eqref{metricMalpha3}, and $\alpha^{\theta\theta}$ represents the contravariant form of $\alpha_{\theta\theta}$ in the metric~\eqref{metricMalpha3}.

Moreover, the Gaussian curvature of $M^{\left(\alpha 2\right)}$ can be obtained by \cite{werner2012gravitational},
\begin{equation}
    K=\frac{1}{\sqrt{\alpha}}\left[\frac{\partial}{\partial \phi}\left(\frac{\sqrt{\alpha}}{\alpha_{r r}} \Gamma_{r r}^\phi\right)-\frac{\partial}{\partial r}\left(\frac{\sqrt{\alpha}}{\alpha_{r r}} \Gamma_{r \phi}^\phi\right)\right].
    \label{GaussianCurvature}
\end{equation}
where all quantities come from metric~\eqref{metricMalpha2}, $\alpha$ and $\Gamma$ are the determinant and Christoffel symbol, respectively.

\subsection{Applying GWOIA method to asymptotically flat spacetimes}
\label{asymptoticallyflatSAS}
\begin{figure}[!ht]
    \centering
        \includegraphics[width=0.6\columnwidth]{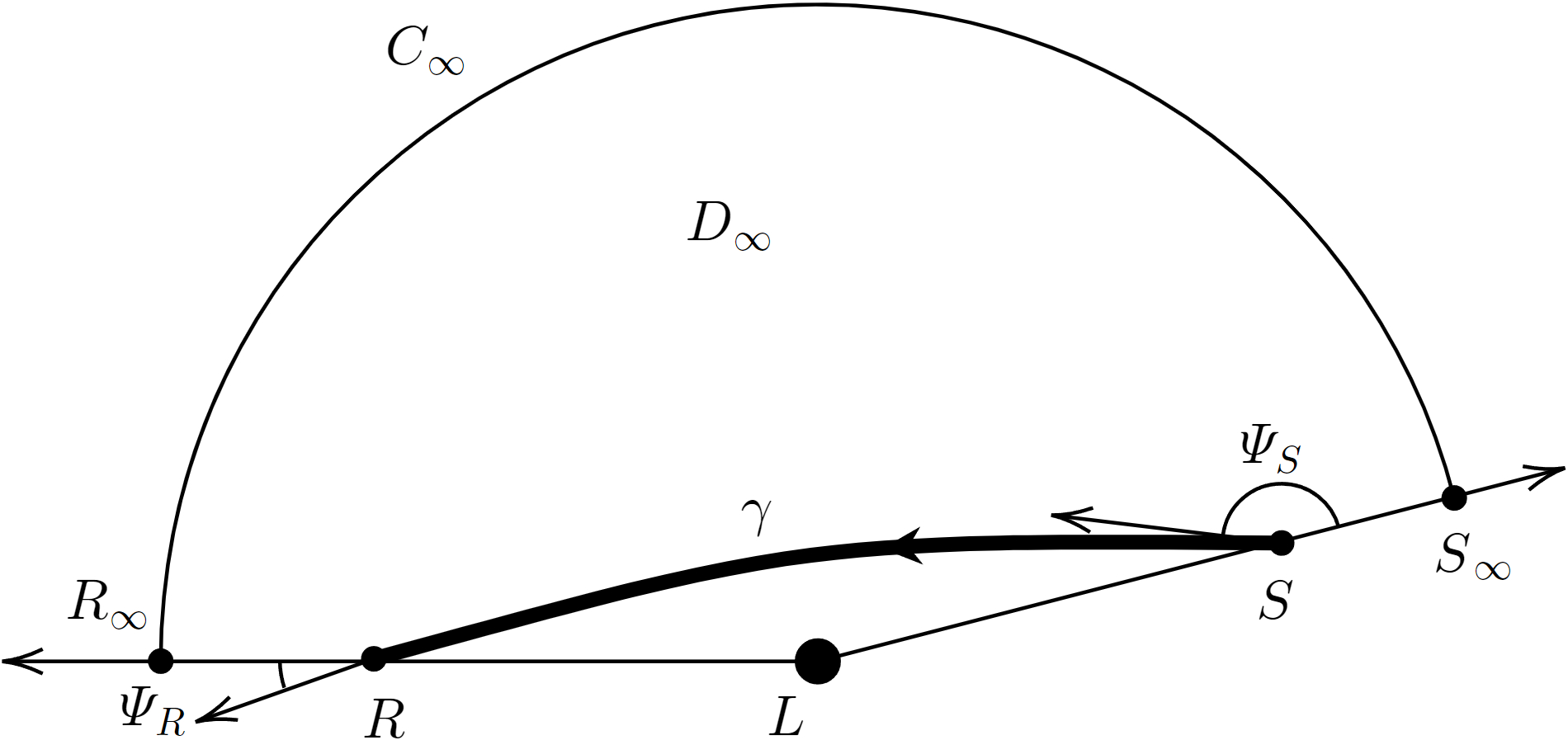} 
    \caption{The quadrilateral region $_{B_\infty}^B \square_{A_\infty}^A$ embedded in the $M^{(\alpha2)}$ corresponding to an SAS spacetime.}
    \label{fig-3}
\end{figure}

As shown in Fig.~\ref{fig-3}, 
in the $M^{\left(\alpha 2\right)}$ corresponding to the SAS spacetime equipped with metric~\eqref{SASmetric}, $L$, $S$, $R$, $\gamma$, $\Psi_S$, and $\Psi_R$ have the same meaning as that in Fig.~\ref{fig-2}, $C_\infty$ is an infinite circular arc, intersecting the outgoing radial curve $\overrightarrow{LS}$ and $\overrightarrow{LR}$ at $S_\infty$ and $R_\infty$, respectively. Then the integral region is constructed as $D_\infty = _{R}^{R_\infty} \square_{S}^{S_\infty}$. Applying the GBT Eq.~\eqref{gbt} to $D_\infty$ leads to
\begin{equation}
  \iint_{D_\infty} K \mathrm{d}S + \int_{\overrightarrow{SS}_\infty} \kappa \mathrm{d}l + \int_{C_\infty} \kappa \mathrm{d}l + \int_{\overrightarrow{R_\infty R}} \kappa \mathrm{d}l  + \int_{\overset{\curvearrowright}{RS}} \kappa \mathrm{d}l + \eta_S + \eta_{S_\infty} + \eta_{R_\infty} + \eta_R = 2\pi \chi\left( D_\infty\right).
\label{gbt01}
\end{equation}
$\int_{\overrightarrow{SS}_\infty} \kappa \mathrm{d}l  = \int_{\overrightarrow{R_\infty R}} \kappa \mathrm{d}l=0$, since $\overrightarrow{SS_\infty}$ and $\overrightarrow{R_\infty R}$ are geodesics (see Appendix A of Ref.~\cite{huang2023finite} for the proof). $\int_{\overset{\curvearrowright}{RS}} \kappa \mathrm{d}l = - \int_{\gamma} \kappa \mathrm{d}l$, $\eta_S=\pi-\Psi_S$, $\eta_R = \Psi_R$, $\eta_{S_\infty}=\eta_{R_\infty}=\pi/2$, and $\chi\left(D_\infty\right)=1$ as $D_\infty$ is simply connected. 

In the scenario where the spacetime is asymptotically flat, $C_\infty$ can be treated as a circular arc in the flat space, i.e. $\int_{C_\infty} \kappa \cdot \mathrm{d}l = \int_{\phi_S}^{\phi_R} \left( 1/r_\infty \right)\cdot r_\infty \mathrm{d}\phi =\phi_{RS}$. According to Eq.~\eqref{gbt01} and the related results, we can express the deflection angle as
\begin{equation}
    \delta = -\iint_{D_\infty}K \mathrm{d}S + \int_{\gamma} \kappa \mathrm{d}l,
    \label{deltaBA0}
\end{equation}
in which the definition Eq.~\eqref{defAngle} is adopted. The above formula has been used to calculate the deflection angle of particles in Refs.~\cite{ono2018deflection,ovgun2018light,ovgun2018shadow,haroon2019shadow,ono2019deflection,ono2019effects,kumar2019shadow,li2020equivalence,li2020thefinitedistance,li2021kerr,li2021kerrnewman,li2022deflection,pantig2023testing}, where the spacetime is asymptotically flat.

\subsection{Applying GWOIA method to asymptotically nonflat spacetimes}
\label{asymptoticallynonflatspacetimes1}
Based on the presence or absence of the singularity as the radial coordinate approaches infinity, asymptotically nonflat SAS spacetimes can be divided into two categories: the infinity-reachable and the infinity-unreachable. Examples of the infinity-reachable include the Kerr-like black hole in bumblebee gravity \cite{ding2019thin,ding2020exact}, while the Kerr-de Sitter spacetime \cite{dewitt1973black} and the rotating solution in conformal Weyl gravity \cite{mannheim1991solutions,varieschi2014kerr} fall under the category of the infinity-unreachable.

In the case of infinity-reachable asymptotically nonflat SAS spacetimes, the construction of an infinite integral region is still allowed. However, the calculation formula of the deflection angle becomes more intricate and relies on the specific metric. To illustrate this, we briefly review the work in \cite{li2020finite}, where the Kerr-like black hole in bumblebee gravity is considered. The metric for such black hole is given by
\begin{equation}
    \mathrm{d} s^2=  -\left(1-\frac{2 M r}{\rho^2}\right) \mathrm{d} t^2-\frac{4 M a r \lambda \sin ^2 \theta}{\rho^2} \mathrm{d} \phi \mathrm{d} t +\frac{\rho^2}{\Delta} \mathrm{d}r^2+\rho^2 \mathrm{d} \theta^2+\frac{A \sin ^2 \theta}{\rho^2} \mathrm{d} \phi^2,
    \label{appKBG}
\end{equation}
in which $\lambda =\sqrt{1+l}$, $\rho^2 =r^2+\lambda^2 a^2 \cos ^2 \theta$, $\Delta =\left(r^2-2 M r\right)/\lambda^2+a^2$, $A =\left(r^2+\lambda^2 a^2\right)^2-\Delta \lambda^4 a^2 \sin ^2 \theta$. Here, $M$ and $a$ are the mass and rotation parameter of the black hole, respectively, and $l$ is the Lorentz violation parameter. In this scenario, Fig.~\ref{fig-3} can still serve as the illustration, and the relevant constructions and results, including Eq.~\eqref{gbt01}, remain valid except for the integral of the geodesic curvature along $C_\infty$. The author derived $\int_{C_\infty}\kappa \mathrm{d}l = \left(1/\lambda\right) \phi_{RS}$ instead of the result obtained in asymptotically flat spacetime. Consequently, the deflection angle is expressed as
\begin{equation}
    \delta = -\iint_{D_\infty}K \mathrm{d}S + \int_{\gamma} \kappa \mathrm{d}l + \left(1-\frac{1}{\lambda}\right)\phi_{RS},
    \label{deltaBA1}
\end{equation}
in which an additional item, the change in the coordinate angle, appears compared to Eq.~\eqref{deltaBA0} due to the existence of the bumblebee vector field. It is important to note that the formula for the deflection angle will vary depending on the specific metrics employed.

As for the infinity-unreachable asymptotically nonflat SAS spacetime, the infinite integral region $D_\infty$ is ill-defined, rendering the GWOIA method invalid.

\section{Generalized GW method}
\label{sec-4}
In this section, we present a generalized GW method to solve the ill-defined problem and simplify the related calculation. Denoting the radial coordinate of the auxiliary circular arc as $r_c$, we give up the limit $r_c\to\infty$ (or $r_c=\infty$) in the conventional GWOIA method and prove that the $r_c$ can be chosen arbitrarily, the only requirement is that the resulting integral region should not contain the physically unreasonable region (the region with singularity points).

The generalized GW method is elaborated by considering three different situations based on the relation between $r_c$ and the maximum ($r_\gamma^{max}$) and the minimum ($r_\gamma^{min}$) of the radial coordinates of trajectories. These situations are $r_c>r_\gamma^{max}$, $r_\gamma^{max}\ge r_c\ge r_\gamma^{min}$, and $r_c<r_\gamma^{min}$. Like the previous section, the $M^{\left(\alpha 2\right)}$ represents the equatorial plane of the Riemannian part of the JMRF metric corresponding to an SAS spacetime equipped with metric~\eqref{SASmetric}, and its metric is expressed as Eq.~\eqref{metricMalpha2}.

\subsection{$r_c>r_\gamma^{max}$}
\begin{figure}[!ht]
    \centering
        \includegraphics[width=0.6\columnwidth]{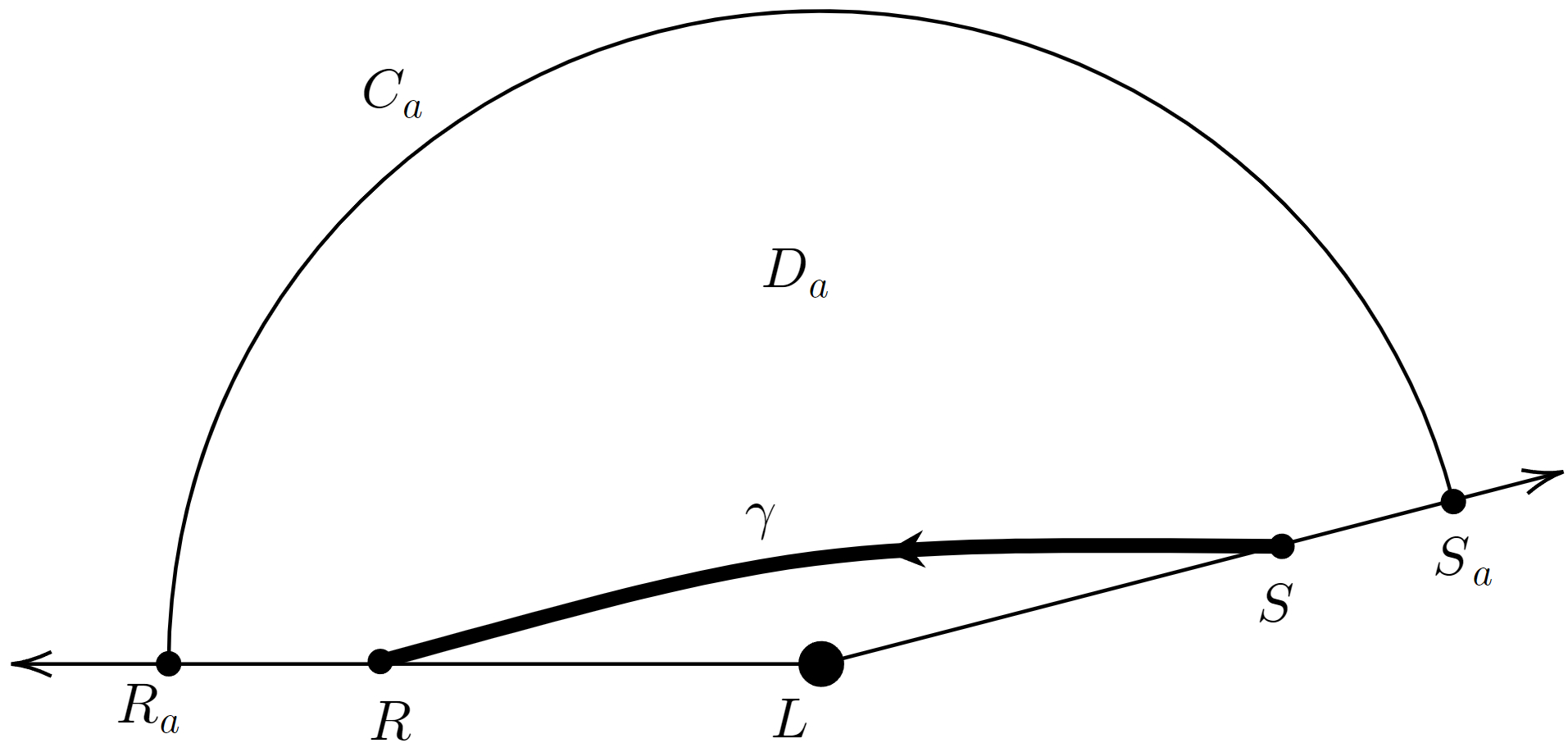} 
    \caption{The schematic for the situation $r_c>r_\gamma^{max}$.}
    \label{fig-4}
\end{figure}

As shown in Fig.~\ref{fig-4}, 
in the $M^{\left(\alpha 2\right)}$, $L$, $S$, $R$, and $\gamma$ have the same meaning as that in Fig.~\ref{fig-2}. $C_a=\overset{\curvearrowright}{S_a R_a}$ is the auxiliary circular arc with $r_c>r_\gamma^{max}$, and intersects with $\overrightarrow{LS}$ and $\overrightarrow{LR}$ at $S_a$ and $R_a$, respectively. Thus we obtain a quadrilateral region $D_a=^{R_a}_R\square^{S_a}_S$. Here, the subscript $a$ indicates "arbitrary". The application of the GBT to $D_a$ yields
\begin{equation}
  \iint_{D_a} K \mathrm{d}S + \int_{\overrightarrow{SS}_a} \kappa \mathrm{d}l + \int_{C_a} \kappa \mathrm{d}l + \int_{\overrightarrow{R_aR}} \kappa \mathrm{d}l + \int_{\overset{\curvearrowright}{RS}} \kappa \mathrm{d}l + \eta_S + \eta_{S_a} + \eta_{R_a} + \eta_R = 2\pi \chi(D_a).
\end{equation}
Substituting $\kappa(\overrightarrow{SS}_a)=\kappa(\overrightarrow{R_aR})=0$, $\int_{\overset{\curvearrowright}{RS}} \kappa \mathrm{d}l=-\int_\gamma \kappa \mathrm{d}l$, $\eta_S = \pi - \Psi_S$, $\eta_{S_a}=\eta_{R_a}=\pi/2$, $\eta_R=\Psi_R$, and $\chi\left(D_a\right)=1$ into the above equation leads to
\begin{equation}
    \iint_{D_a} K \mathrm{d}S +\int_{C_a} \kappa \mathrm{d}l - \int_\gamma \kappa \mathrm{d}l +\Psi_R - \Psi_S = 0.
\end{equation}
With the definition Eq.~\eqref{defAngle}, the deflection angle can be written as
\begin{equation}
    \delta = - \iint_{D_a} K \mathrm{d}S -\int_{C_a} \kappa \mathrm{d}l   +\phi_{RS}+ \int_\gamma \kappa \mathrm{d}l.
    \label{deltageneral}
\end{equation}
Firstly, we analyze $\iint_{D_a} K \mathrm{d}S$. According to Eqs.~\eqref{metricMalpha2} and \eqref{GaussianCurvature}, we have
\begin{equation}
        \int K\sqrt{\alpha} \mathrm{d}r = - \frac{\alpha_{\phi\phi,r}}{2\sqrt{\alpha}} + Const.
    \label{rintegral}
\end{equation}
Introducing $H(r)$ to denote the indefinite integral of $K\sqrt{\alpha}$ with respect to the radial coordinate up to a constant, namely,
\begin{equation}
    H(r)=-\frac{\alpha_{\phi\phi,r}}{2\sqrt{\alpha}},
    \label{Hr}
\end{equation}
we derive
\begin{equation}
    \iint_{D_a} K \mathrm{d}S = \int_{\phi_S}^{\phi_R} \int_{r_\gamma}^{r_c} K\sqrt{\alpha} \mathrm{d}r \mathrm{d}\phi = \int_{\phi_S}^{\phi_R} \left[ H(r_c) - H(r_\gamma)  \right] \mathrm{d}\phi.
    \label{gaosiqulvDa}
\end{equation}
Secondly, we analyze $\int_{C_a} \kappa \mathrm{d}l$. According to Liouville's formula for geodesic curvature, the geodesic curvature of the circular arc in $M^{(\alpha 2)}$ can be expressed as (Chapter 4 of Ref.~\cite{struik1961lectures})
\begin{equation}
    \kappa^{(c)} = -\Gamma^r_{\phi\phi}\frac{\alpha^{1/2}}{\alpha^{3/2}_{\phi\phi}}  = \frac{\alpha_{\phi\phi,r}}{2 \alpha_{\phi\phi}\sqrt{\alpha_{rr}}}.
    \label{cediqulvcircular}
\end{equation}
Introducing $G(r)=\kappa^{(c)} \mathrm{d}l/\mathrm{d}\phi$ to denote the integrand of the integral of geodesic curvature along the auxiliary circular arc with respect the azimuthal coordinate, we derive
\begin{equation}
    \int_{C_a}\kappa \mathrm{d}l = \int_{\phi_S}^{\phi_R} \left[ \kappa^{(c)} \frac{\mathrm{d}l}{\mathrm{d}\phi} \right]_{r=r_c} \mathrm{d}\phi = \int_{\phi_S}^{\phi_R} G(r_c) \mathrm{d}\phi.
     \label{kappajifen}
\end{equation}
According to Eqs.~\eqref{metricMalpha2} and \eqref{cediqulvcircular} we obtain
\begin{equation}
    G(r)  = \frac{\alpha_{\phi\phi,r}}{2\sqrt{\alpha}}.
    \label{Gr}
\end{equation}
Finally, substituting Eqs.~\eqref{gaosiqulvDa} and \eqref{kappajifen} into Eq.~\eqref{deltageneral} yields
\begin{equation}
    \delta =  -\int_{\phi_S}^{\phi_R} \left[ H(r_c) - H(r_\gamma) +G(r_c) - 1  \right] \mathrm{d}\phi + \int_\gamma \kappa \mathrm{d}l.
    \label{generaldelta}
\end{equation}
We find that $H(r)=-G(r)$ regardless of the value of $r$ according to Eqs.~\eqref{Hr} and \eqref{Gr}, thus the above formula becomes
\begin{equation}
    \delta =   \int_{\phi_S}^{\phi_R} \left[1+ H(r_\gamma)  \right] \mathrm{d}\phi + \int_\gamma \kappa \mathrm{d}l.
    \label{GGMdelta}
\end{equation}

\subsection{$r_\gamma^{max}\ge r_c\ge r_\gamma^{min}$}
\begin{figure}[!ht]
    \centering
        \includegraphics[width=0.6\columnwidth]{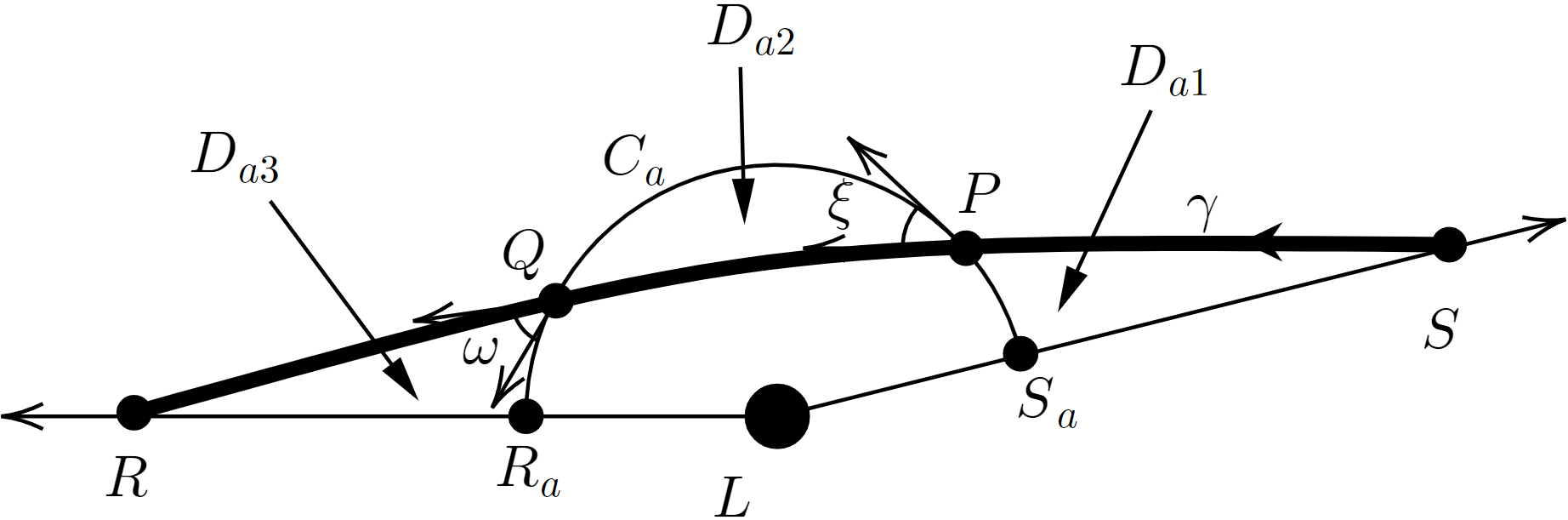} 
    \caption{The schematic for the situation $r_\gamma^{max}\ge r_c\ge r_\gamma^{min}$.}
    \label{fig-5}
\end{figure}

Now, we show that Eq.~\eqref{GGMdelta} also holds for the situation $r_\gamma^{max}\ge r_c\ge r_\gamma^{min}$. As depicted in Fig.~\ref{fig-5}, 
the auxiliary circular arc $C_a$ satisfies $r_\gamma^{max}\ge r_c\ge r_\gamma^{min}$, and intersects with $\gamma$ at $P$ and $Q$. Thus we obtain two triangle regions $D_{a1}=_{P}\triangle_{S_a}^S$ and $D_{a3}=_{R}\triangle^{Q}_{R_a}$, and a digon $D_{a2}$ with vertexes $P$ and $Q$. $\xi$ and $\omega$ are the angle between $C_a$ and $\gamma$ at $P$ and $Q$, respectively.

Applying the GBT to $D_{a1}$ leads to
\begin{equation}
        \iint_{D_{a1}} K \mathrm{d}S + \int_{\overrightarrow{S_aS}} \kappa \mathrm{d}l + \int_{\overset{\curvearrowright}{SP}} \kappa \mathrm{d}l + \int_{\overset{\curvearrowright}{PS_a}} \kappa \mathrm{d}l   + \eta_{S_a} + \eta_S + \eta_P^{(1)}= 2\pi \chi(D_{a1}).
\end{equation}
Using $\kappa(\overrightarrow{S_aS})=0$, $\eta_{S_a}=\pi/2$, $\eta_S=\Psi_S$, $\eta_P^{(1)}=\pi-\xi$, and $\chi(D_{a1})=1$, we derive
\begin{equation}
    \iint_{D_{a1}} K \mathrm{d}S  + \int_{\overset{\curvearrowright}{SP}} \kappa \mathrm{d}l + \int_{\overset{\curvearrowright}{P S_a}} \kappa \mathrm{d}l + \Psi_S -\xi = \frac{\pi}{2} .\label{rst1}
\end{equation}
Applying the GBT to $D_{a2}$ leads to
\begin{equation}
   \iint_{D_{a2}} K \mathrm{d}S + \int_{\overset{\curvearrowright}{QP}} \kappa \mathrm{d}l+ \int_{\overset{\Large\curvearrowright}{PQ}}  \kappa \mathrm{d}l + \eta_P^{\left(2\right)}   + \eta_Q^{\left(2\right)} = 2\pi \chi\left(D_{a2}\right),
\end{equation}
where $\overset{\curvearrowright}{QP}$ denotes the curve followed from $Q$ to $P$ along the trajectory $\gamma$, $\overset{\curvearrowright}{PQ}$ represents the curve followed from $P$ to $Q$ along the circular arc $C_a$. Using $\eta_P^{\left(2\right)}=\pi-\xi$, $\eta_Q^{\left(2\right)}=\pi-\omega$, and $\chi(D_{a2})=1$, we derive
\begin{equation}
    \iint_{D_{a2}} K \mathrm{d}S + \int_{\overset{\curvearrowright}{QP}} \kappa \mathrm{d}l+ \int_{\overset{\Large\curvearrowright}{PQ}}  \kappa \mathrm{d}l -\xi-\omega = 0. \label{rst2}
\end{equation}
Applying the GBT to $D_{a3}$ leads to
\begin{equation}
   \iint_{D_{a3}} K \mathrm{d}S + \int_{\overset{\curvearrowright}{QR}} \kappa \mathrm{d}l + \int_{\overrightarrow{RR_a}} \kappa \mathrm{d}l + \int_{\overset{\curvearrowright}{R_aQ}} \kappa \mathrm{d}l  + \eta_Q^{(3)} + \eta_{R}+ \eta_{R_a}= 2\pi \chi\left(D_{a3}\right).
\end{equation}
Using $\kappa\left(\overrightarrow{RR_a}\right)=0$, $\eta_Q^{(3)} = \pi-\omega$, $\eta_{R}=\pi-\Psi_R$, $\eta_{R_a}=\pi/2$, and $\chi\left(D_{a3}\right)=1$, we derive
\begin{equation}
    \iint_{D_{a3}} K \mathrm{d}S + \int_{\overset{\curvearrowright}{QR}} \kappa \mathrm{d}l+ \int_{\overset{\curvearrowright}{R_aQ}} \kappa \mathrm{d}l  -\omega-\Psi_R=-\frac{\pi}{2}. \label{rst3}
\end{equation}
Acoording to Eq.~\eqref{defAngle}, Eq.~\eqref{rst1}$+$Eq.~\eqref{rst3}$-$Eq.~\eqref{rst2} results in
\begin{equation}
    \delta = \iint_{D_{a1}} K \mathrm{d}S -\iint_{D_{a2}} K \mathrm{d}S +\iint_{D_{a3}} K \mathrm{d}S  - \int_{C_a} \kappa \mathrm{d}l+\phi_{RS}+ \int_\gamma \kappa \mathrm{d}l .
\end{equation}
With the help of Eqs.~\eqref{Hr} and \eqref{Gr}, the deflection angle is expressed as
\begin{equation}
    \begin{aligned}
        \delta = & \int_{\phi_S}^{\phi_P} \left[ H(r_\gamma) - H(r_c)\right] \mathrm{d}\phi - \int_{\phi_P}^{\phi_Q} \left[ H(r_c)-H(r_\gamma)\right] \mathrm{d}\phi  + \int_{\phi_Q}^{\phi_R} \left[ H(r_\gamma) - H(r_c)\right] \mathrm{d}\phi \\
        &  - \int_{\phi_S}^{\phi_R} G(r_c) \mathrm{d}\phi +\phi_{RS} + \int_\gamma \kappa \mathrm{d}l \\
        =& \int_{\phi_S}^{\phi_R} \left[ H(r_\gamma) - H(r_c)\right] \mathrm{d}\phi - \int_{\phi_S}^{\phi_R} G(r_c) \mathrm{d}\phi +\phi_{RS} +  \int_\gamma \kappa \mathrm{d}l  \\
        =&  \int_{\phi_S}^{\phi_R} \left[1+ H(r_\gamma)  \right] \mathrm{d}\phi + \int_\gamma \kappa \mathrm{d}l,
    \end{aligned}
\end{equation}
which is the same as Eq.~\eqref{GGMdelta}. Additionally, if $C_a$ intersects with $\gamma$ at more or fewer points, one can also obtain the above formula by applying the GBT to more or fewer quadrilateral, triangle, or digon regions.

\subsection{$r_c<r_\gamma^{min}$}
\begin{figure}[!ht]
    \centering
        \includegraphics[width=0.6\columnwidth]{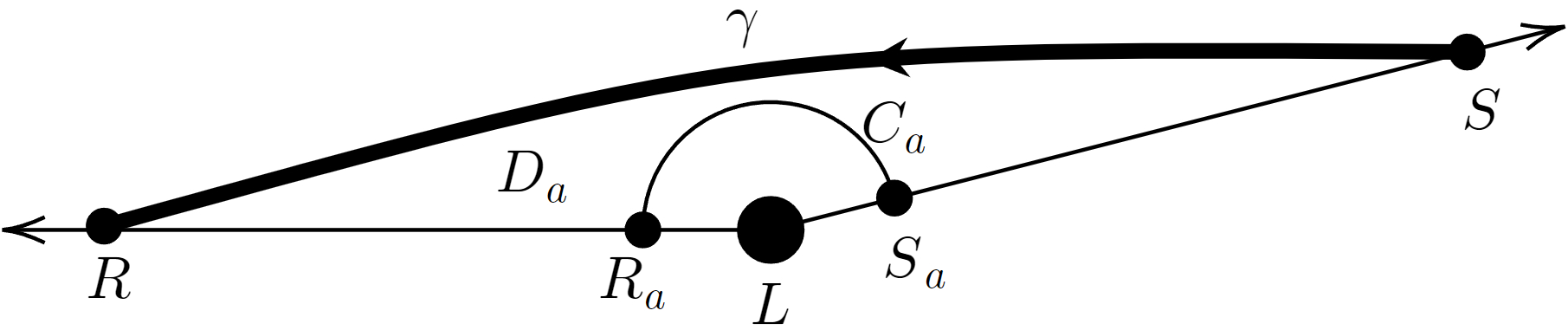} 
    \caption{The schematic for the situation $r_c<r_\gamma^{min}$.}
    \label{fig-6}
\end{figure}

As shown in Fig.~\ref{fig-6}, 
$r_c<r_\gamma^{min}$, applying the GBT to $D_a=^R_{R_a}\square_{S_a}^S$ yields
\begin{equation}
   \iint_{D_a} K \mathrm{d}S + \int_{\overrightarrow{S_aS}} \kappa \mathrm{d}l + \int_{\overset{\curvearrowright}{SR}} \kappa \mathrm{d}l + \int_{\overrightarrow{RR}_a} \kappa \mathrm{d}l  + \int_{\overset{\curvearrowright}{R_aS_a}} \kappa \mathrm{d}l + \eta_{S_a} + \eta_S + \eta_R + \eta_{R_a} = 2\pi \chi(D_a).
\end{equation}
Using $\kappa(\overrightarrow{S_aS})=\kappa(\overrightarrow{RR}_a)=0$, $\eta_{S_a}=\eta_{R_a}=\pi/2$, $\eta_S=\Psi_S$, $\eta_R=\pi-\Psi_R$, $\chi\left(D_a\right)=1$ and Eq.~\eqref{defAngle}, we derive
\begin{equation}
    \delta  = \iint_{D_a} K \mathrm{d}S + \int_{\overset{\curvearrowright}{R_aS_a}} \kappa \mathrm{d}l  + \phi_{RS} + \int_\gamma \kappa \mathrm{d}l.
    \label{toLi}
\end{equation}
Namely,
\begin{equation}
\begin{aligned}
    \delta  = & \int_{\phi_S}^{\phi_R} \left[ H(r_\gamma) - H(r_c) \right] \mathrm{d}\phi - \int_{\phi_S}^{\phi_R} G(r_c) \mathrm{d}\phi  + \phi_{RS} + \int_\gamma \kappa \mathrm{d}l \\
     =&  \int_{\phi_S}^{\phi_R} \left[1+ H(r_\gamma) \right] \mathrm{d}\phi+ \int_\gamma \kappa \mathrm{d}l,
\end{aligned}
\end{equation}
which is the same as Eq.~\eqref{GGMdelta}.

\subsection{Further simplifying the calculation formula and briefly summarizing the calculation step}
\label{calpro}
Although Eq.~\eqref{GGMdelta} is simpler than that in the conventional GWOIA method (Eqs.~\eqref{deltaBA0} and \eqref{deltaBA1}), calculating the integral of geodesic curvature along the trajectory is still not straightforward. Here, we recast it into a form that further simplifies the calculation process for practical computations. We denote the integrand of the integral of geodesic curvature along the trajectory with respect to the azimuthal coordinate by $T\left(r\right)=\kappa_\gamma \mathrm{d}l/\mathrm{d}\phi$. Then using Eqs.~\eqref{metricMalpha3}, \eqref{metricMalpha2} and \eqref{cediqulvcal}, we obtain
\begin{equation}
  T\left(r\right) 
  = -\beta_{\phi,r} \sqrt{ \frac{1}{\alpha_{\phi\phi}} \left( \frac{\mathrm{d}r}{\mathrm{d}\phi}\right)^2+ \frac{1}{\alpha_{rr}} } ,
  \qquad \text{or}\qquad
  T\left(r\right) = -\beta_{\phi,r} \sqrt{ \frac{r^4}{\alpha_{\phi\phi}} \left( \frac{\mathrm{d}u}{\mathrm{d}\phi}\right)^2+ \frac{1}{\alpha_{rr}} }.
  \label{Tr}
\end{equation}
As a consequence, we express the integral of geodesic curvature along $\gamma$ as
\begin{equation}
    \int_\gamma \kappa \mathrm{d}l = \int_{\phi_S}^{\phi_R} \left. \left[ \kappa_\gamma \frac{\mathrm{d}l}{\mathrm{d}\phi}\right]\right|_{r=r_\gamma} \mathrm{d}\phi = \int_{\phi_S}^{\phi_R} T(r_\gamma) \mathrm{d}\phi,
\end{equation}
with which Eq.~\eqref{GGMdelta} is finally simplified as
\begin{equation}
    \delta = \int_{\phi_S}^{\phi_R} \left[1+ H(r_\gamma)+T(r_\gamma) \right] \mathrm{d}\phi.
    \label{delta19}
\end{equation}
This formula is applicable to all kinds of SAS spacetime, including the asymptotically flat, the infinity-reachable asymptotically nonflat, and the infinity-unreachable asymptotically nonflat. It provides an unified description for the finite-distance deflection angle of particles moving in the equatorial plane of SAS spacetimes, the calculation process using the simplified formula Eq.~\eqref{delta19} is simpler and more straightforward. It can be summarized as follows: 
\begin{itemize}
    \item Firstly, substitute the metric into Eq.~\eqref{dudphi} to obtain the $\left(\mathrm{d}u/\mathrm{d}\phi\right)^2$, and the orbit solution $u\left(\phi\right)=1/r(\phi)$ and $\phi\left(u\right)$.
    \item Secondly, substitute the metric into Eqs.~\eqref{alpharralphaphiphi} and \eqref{betaMalpha2} to get $\alpha_{rr}$, $\alpha_{\phi\phi}$ and $\beta_\phi$.
    \item Thirdly, substitute $\alpha_{rr}$, $\alpha_{\phi\phi}$, $\beta_\phi$, $\left(\mathrm{d}u/\mathrm{d}\phi\right)^2$ and $r_\gamma=1/u(\phi)$ into the integrand of Eq.~\eqref{delta19} to derive $f(\phi)$, where
    \begin{equation}
        f\left(\phi\right)=1+ H(r_\gamma)+T(r_\gamma).
        \label{fphi}
    \end{equation}
    \item Finally, obtain the deflection angle in terms of $u_R$ and $u_S$ with
    \begin{equation}
        \delta = F\left(\phi_R\right)-F\left(\phi_S\right),
        \label{deltaFF}
    \end{equation}
    where $\phi_R=\phi(u_R)$ and $\phi_S=\phi(u_S)$, $\phi_R>\pi/2$ and $\phi_S<\pi/2$ are usually assumed, $u_R$ and $u_S$ are respectively the reciprocal of the radial coordinate of the observer (receiver) and source, and $F\left(\phi\right)$ denotes the indefinite integral of $f\left(\phi\right)$. 
\end{itemize}

The scheme of the generalized GW encompasses various special cases, such as the deflection angle of photons ($v=1$), the infinite-distance deflection angle ($u_S=u_R=0$, only valid for spacetimes where the source and observer can reach infinity), and the deflection angle of SSS counterparts (the rotation parameter vanishes). Furthermore, for the GW method, numerous examples demonstrate that if the spacetime can be approximated as Minkowski spacetime ($\mathrm{d}s^2=-\mathrm{d}t^2+\mathrm{d}r^2+r^2\mathrm{d}\theta^2+r^2\sin^2\theta\mathrm{d}\phi^2$) in terms of the quantities of interest under the zeroth-order approximation, the $n$-th order deflection angle can be obtained using the $(n-1)$-th order orbit solution.

\subsection{Some discussion}
\label{somediscussion}
Our discovery of the expression $H(r_c)=-G(r_c)$ is the cornerstone of the generalized GW method. It neutralizes the $G(r_c)$ in the integral of geodesic curvature and the $H(r_c)$ in the integral of Gaussian curvature regardless of the chosen value of $r_c$. The resulting Eq.~\eqref{delta19} does not depend on the location of the auxiliary circular arc. Our method encompasses the conventional GWOIA method as a specific case with $r_c=\infty$. To illustrate the relationship between the two methods, we employ the generalized GW method with $r_c=\infty$ to reproduce the calculation formula in the conventional GWOIA method. When $r_c=\infty$ the deflection angle can be expressed by Eq.~\eqref{generaldelta} since $\infty>r_\gamma^{max}$. Firstly, we reproduce Eq.~\eqref{deltaBA0}, which corresponds to asymptotically flat SAS spacetimes. Given the asymptotic flatness of spacetimes, we have $\alpha_{rr}\left(\infty\right)=1$ and $\alpha_{\phi\phi}\left(\infty\right)=r^2$ according to Eq.~\eqref{metricMalpha2}. Substituting these values into Eq.~\eqref{Gr} yields $G\left(r_c=\infty\right) = 1$, with which Eq.~\eqref{generaldelta} can reduce to Eq.~\eqref{deltaBA0}. Secondly we reproduce Eq.~\eqref{deltaBA1}, which corresponds to an infinity-reachable nonflat SAS spacetime (Kerr-like black hole in bumblebee gravity). Using the result $G(r_c=\infty)=1/\lambda$ from Ref.~\cite{li2020finite}, Eq.~\eqref{generaldelta} directly reduces to Eq.~\eqref{deltaBA1}. 

The generalized GW and the corresponding simplified calculation formula holds significant meaning in five aspects: 
\begin{itemize}
\item[(a)] The GW method can be extended to infinity-unreachable spacetimes by constructing an appropriate finite integral region, since the auxiliary circular arc can be chosen arbitrarily instead of taking the limit $r_c=\infty$ in previous works.
\item[(b)] For infinity-reachable asymptotically nonflat SAS spacetimes, our calculation formula Eq.~\eqref{delta19} is universally applicable, unlike that in the conventional GWOIA method which varies depending on specific metrics.
\item[(c)] Our method can be viewed as an unified description to the GW method for calculating the finite-distance and infinite-distance deflection angle of massive and massless particles in SSS and SAS spacetimes with or without asymptotical flatness. This is the reason why we call it the generalized GW method.
\item[(d)] Compared to calculation formulas in the conventional GWOIA method (Eqs.~\eqref{deltaBA0} and \eqref{deltaBA1}), our formula Eq.~\eqref{delta19} is easier to apply in practical computation processes. Since it involves only a single integral of an fully simplified integrand, rather than a double integral of the intricate Gaussian curvature and a single integral of the geodesic curvature required by Eqs.~\eqref{deltaBA0} and \eqref{deltaBA1}.
\item[(e)] Since the presence of the surface integral over $D_\infty$ in the calculation formula in conventional GWOIA method (Eqs.~\eqref{deltaBA0} and \eqref{deltaBA1}), the deflection angle seems depend on the nature of the region between the trajectory and infinity. While our result, Eq.~\eqref{delta19}, clearly demonstrates that the deflection angle is solely determined by the properties of the trajectory itself, which is closer to our intuitive understanding.
\end{itemize}

\section{Application of Generalized GW method}
\label{sec-5}
In this section, we aim to showcase the calculation process outlined in Sec.\ref{calpro} and demonstrate the conclusion drawn in Sec.~\ref{somediscussion}. To this end, by utilizing the generalized GW method along with the simplified formula Eq.~\eqref{delta19}, we compute the finite-distance deflection angle of particles for three spacetimes: an asymptotically flat SAS spacetime, Kerr black hole; an infinity-reachable asymptotically nonflat SAS spacetime, Kerr-like black hole in bumblebee gravity; and an infinity-unreachable asymptotically nonflat SAS spacetime, rotating solution in conformal Weyl gravity.

\subsection{Asymptotically flat SAS spacetime (Kerr black hole)}
The Kerr metric with Boyer-Lindquist coordinates states \cite{boyer1967maximal}
\begin{equation}
      \mathrm{d} s^{2}= -\left(1-\frac{2 M r}{\Sigma}\right) \mathrm{d} t^{2}-\frac{4 a M r \sin ^{2} \theta}{\Sigma} \mathrm{d} t \mathrm{d} \phi +\frac{\Sigma}{\Delta} \mathrm{d} r^{2}+\Sigma \mathrm{d} \theta^{2}+\left[\Delta+\frac{2Mr\left(r^2+a^2\right)}{\Sigma}\right] \sin ^{2} \theta \mathrm{d} \phi^{2},
      \label{KerrMetric}
\end{equation}
where $\Sigma = r^{2}+a^{2} \cos ^{2} \theta$, $\Delta=  r^{2}-2 M r+a^{2}$. We adopt the calculation steps in Sec.\ref{calpro} to calculate the deflection angle of particles moving in the equatorial plane of Kerr spacetime. Firstly, according to Eq.~\eqref{dudphi}, we obtain the equation of motion
\begin{equation}
    \left(\frac{\mathrm{d}u}{\mathrm{d}\phi}\right)^2= \frac{1-b^2 u^2}{b^2}+ M\cdot \frac{ 2 u \left(b^2 u^2 v^2-v^2+1\right)}{b^2 v^2}  + \mathcal{O}\left(M^2, Ma, a^2\right).
\label{Kerrdudphi2}
\end{equation}
Then the orbit solution can be derived with the perturbation method \footnote{When $M=a=0$, Kerr metric becomes the Minkowski spacetime. Here we calculate the second order deflection angle with respect to $M$ and $a$, thus the orbit solution up to first order is enough.}
\begin{equation}
    u\left(\phi\right) = \frac{\sin\phi}{b}+M\cdot \frac{1+v^2 \cos^2\phi}{b^2v^2}+ \mathcal{O}\left(M^2, Ma, a^2\right).
    \label{uofphi}
\end{equation}
In addition, the iterative solution of $\phi$ can be obtained by using the above formula
\begin{equation}
    \phi(u) = \begin{cases}
    \Phi(u), & \text{if  } \left| \phi \right| <\frac{\pi}{2}, \\
    \pi - \Phi(u) , & \text{if  } \left| \phi \right| >\frac{\pi}{2},
        \end{cases}
        \label{phigamma}
\end{equation}
where
\begin{equation}
    \Phi\left(u\right) = \arcsin\left(bu\right)  -M \cdot \frac{1+v^2-b^2u^2v^2}{bv^2\sqrt{1-b^2u^2}}  + \mathcal{O}\left(M^2, Ma, a^2\right).
\end{equation}
Secondly, we substitute the metric \eqref{KerrMetric} into Eqs.~\eqref{alpharralphaphiphi} and \eqref{betaMalpha2} to obtain $\alpha_{rr}$, $\alpha_{\phi\phi}$, and $\beta_\phi$. Thirdly, substituting $\alpha_{rr}$, $\alpha_{\phi\phi}$, $\beta_\phi$, and Eq.~\eqref{Kerrdudphi2} into Eqs.~\eqref{Hr} and \eqref{Tr}, and using Eq.~\eqref{fphi}, we derive
\begin{equation}
    \begin{aligned}
        f(\phi)
        = & M\cdot \frac{\left( v^{2} +1\right)\sin \phi }{bv^{2}} +M^{2} \cdot \frac{1}{4b^{2} v^{4}} \left[\left( v^{2} -2\right)^{2}\cos (2\phi )+3\left( v^{2} +4\right) v^{2}\right]\\
         & -Ma\cdot \frac{2\sin \phi }{b^{2} v} +\mathcal{O}\left( M^{3} ,M^{2} a,Ma^{2} ,a^{3}\right),
        \end{aligned}
\end{equation}
in which $r_\gamma$ is expressed as the reciprocal of Eq.~\eqref{uofphi}. Finally, substituting $F\left(\phi\right)$ (derived by integrating $f(\phi)$), $\phi_R=\pi-\Phi\left(u_R\right)$, and $\phi_S=\Phi\left(u_S\right)$ into Eq.~\eqref{deltaFF} yields the deflection angle of particles moving in the equatorial plane
\begin{equation}
    \begin{aligned}
        \delta = & M\cdot \frac{\left( v^{2} +1\right)\left(\sqrt{1-b^{2} u{_{R}}^{2}} +\sqrt{1-b^{2} u{_{S}}^{2}}\right)}{bv^{2}} +  M^{2} \cdot \left\{\frac{3\left( v^{2} +4\right)[\arccos (bu_{R} )+\arccos (bu_{S} )]}{4b^{2} v^{2}}\right. \\
         & +\frac{u_{S}\left[ 3v^{2}\left( v^{2} +4\right) -b^{2} u{_{S}}^{2}\left( 3v^{4} +8v^{2} -4\right)\right]}{4bv^{4}\sqrt{1-b^{2} u{_{S}}^{2}}}\left. +\frac{u_{R}\left[ 3v^{2}\left( v^{2} +4\right) -b^{2} u{_{R}}^{2}\left( 3v^{4} +8v^{2} -4\right)\right]}{4bv^{4}\sqrt{1-b^{2} u{_{R}}^{2}}}\right\}\\
         & -Ma\cdot \frac{2\left(\sqrt{1-b^{2} u_{R}^{2}} +\sqrt{1-b^{2} u_{S}^{2}}\right)}{b^{2} v} +\mathcal{O}\left( M^{3} ,M^{2} a,Ma^{2} ,a^{3}\right).
        \end{aligned}
        \label{daKerr}
\end{equation}
Li and Jia also obtained Eq.~\eqref{daKerr} with the conventional GWOIA method in Ref.~\cite{li2020thefinitedistance}, where the calculation process is much more complicated than ours and the redundant second order term $Ma$ is retained in the orbit solution.

\subsection{Infinity-reachable asymptotically nonflat SAS spacetime (Kerr-like black hole in bumblebee gravity)}
The metric of the Kerr-like black hole in bumblebee gravity is Eq.~\eqref{appKBG}. We calculate the deflection angle of particles moving in the equatorial plane of such spacetime. Firstly, substituting metric~\eqref{appKBG} into Eq.~\eqref{dudphi} leads to

\begin{equation}
        \left(\frac{\mathrm{d}u}{\mathrm{d}\phi }\right)^{2} =  \frac{1-b^{2} u^{2}}{b^{2} \lambda ^{2}} +M\cdot \frac{2\left( b^{2} u^{3} v^{2} -uv^{2} +u\right)}{b^{2} \lambda ^{2} v^{2}} -Ma\cdot \frac{4u}{b^{3} \lambda v} +a^{2} \cdot \frac{3u^{2} -2b^{2} u^{4}}{b^{2}}+ \mathcal{O}\left(M^3,M^2a,Ma^2,a^3\right),
        \label{bumblebeedudphi2}
\end{equation}
with which we obtain the orbit solution of particles moving in the equatorial plane
\begin{equation}
    \begin{aligned}
        u(\phi )= & \frac{1}{b}\sin\frac{\phi }{\lambda } +M\cdot \frac{1}{b^{2}}\left[\cos^{2}\frac{\phi }{\lambda } +\frac{1}{v^{2}}\right] -M^{2} \cdot \frac{1}{8b^{3} \lambda v^{4}}\cos\frac{\phi }{\lambda }\left[ 3\lambda v^{4}\sin\frac{2\phi }{\lambda } +6v^{2}\left( v^{2} +4\right) \phi \right. \\
         & \left. -4\lambda \left( 4v^{2} +1\right)\tan\frac{\phi }{\lambda }\right] -Ma\cdot \frac{2\lambda }{b^{3} v} +a^{2} \cdot \frac{\lambda ^{2}}{2b^{3}}\sin^{3}\frac{\phi }{\lambda } +\mathcal{O}\left( M^{3} ,M^{2} a,Ma^{2} ,a^{3}\right),
        \end{aligned}
        \label{uofphiBumblebee}
\end{equation}
and
\begin{equation}
    \phi(u) = \begin{cases}
    \Phi(u), & \text{if  } \left| \phi \right| <\frac{\pi}{2}, \\
    \lambda\pi - \Phi(u) , & \text{if  } \left| \phi \right| >\frac{\pi}{2},
        \end{cases}
        \label{phiofubumblebee}
\end{equation}
where
\begin{equation}
    \begin{aligned}
        \Phi (u)= & \lambda \arcsin (bu)+M\cdot \frac{\lambda \left( b^{2} u^{2} v^{2} -v^{2} -1\right)}{bv^{2}\sqrt{1-b^{2} u^{2}}} -M^{2} \cdot \frac{\lambda }{4b^{2} v^{4}\left( 1-b^{2} u^{2}\right)^{3/2}} \Big[ 3b^{5} u^{5} v^{4}  \\
        & +3buv^{2}\left( v^{2} +4\right) -3v^{2}\left( v^{2} +4\right)\left( 1-b^{2} u^{2}\right)^{3/2}\arcsin (bu)-2b^{3} u^{3}\left( 3v^{4} +6v^{2} +1\right)\Big]\\
         & +Ma\cdot \frac{2\lambda ^{2}}{b^{2} v\sqrt{1-b^{2} u^{2}}} -a^{2} \cdot \frac{b\lambda ^{3} u^{3}}{2\sqrt{1-b^{2} u^{2}}} +\mathcal{O}\left( M^{3} ,M^{2} a,Ma^{2} ,a^{3}\right).
        \end{aligned}
\end{equation}
Secondly, substituting metric~\eqref{appKBG} into Eqs.~\eqref{alpharralphaphiphi} and \eqref{betaMalpha2} yields the corresponding $\alpha_{rr}$, $\alpha_{\phi\phi}$, and $\beta_\phi$. Thirdly, substituting $\alpha_{rr}$, $\alpha_{\phi\phi}$, $\beta_\phi$, and Eq.~\eqref{bumblebeedudphi2} into Eqs.~\eqref{Hr} and \eqref{Tr}, and using Eq.~\eqref{fphi}, we derive
\begin{equation}
    \begin{aligned}
        f(\phi )= & \frac{\lambda -1}{\lambda } +M\cdot \frac{v^{2} +1}{b\lambda v^{2}}\sin\frac{\phi }{\lambda } -Ma\cdot \frac{2}{b^{2} v}\sin\frac{\phi }{\lambda }\\
         & +M^{2} \cdot \frac{1}{4b^{2} \lambda v^{4}}  \left[\left( v^{2} -2\right)^{2}\cos\frac{2\phi }{\lambda } +3\left( v^{2} +4\right) v^{2}\right] +\mathcal{O}\left( M^{3} ,M^{2} a,Ma^{2} ,a^{3}\right),
        \end{aligned}
\end{equation}
where $r_\gamma$ is expressed as the reciprocal of Eq.~\eqref{uofphiBumblebee}. Finally, substituting $F\left(\phi\right)$ (derived by integrating $f(\phi)$), $\phi_R=\lambda \pi-\Phi\left(u_R\right)$, and $\phi_S=\Phi\left(u_S\right)$ into Eq.~\eqref{deltaFF} yields the deflection angle of particles moving in the equatorial plane
\begin{equation}
    \begin{aligned}
        \delta = & (\lambda -1)[\pi -\arcsin (bu_{R} )-\arcsin (bu_{S} )] \\
        & +M\cdot \left[\frac{\lambda v^{2}\left( 1-b^{2} u_{R}^{2}\right) +\lambda -b^{2} u_{R}^{2}}{bv^{2}\sqrt{1-b^{2} u_{R}^{2}}} +\frac{\lambda v^{2}\left( 1-b^{2} u_{S}^{2}\right) +\lambda -b^{2} u_{S}^{2}}{bv^{2}\sqrt{1-b^{2} u_{S}^{2}}}\right] \\
         & +M^{2} \cdot \frac{1}{4b^{2} v^{4}}\Biggl\{3v^{2} \cdot \left( v^{2} +4\right)\bigl( \pi -\lambda [\arcsin (bu_{R} )+\arcsin (bu_{S} )]\bigr)\\
         & +bu_{R}\Biggl[\sqrt{1-b^{2} u_{R}^{2}}\left( 3\lambda v^{4} +8v^{2} -4\right) +\frac{2b^{2} (1-\lambda )u_{R}^{2}}{\left( 1-b^{2} u_{R}^{2}\right)^{3/2}} +\frac{4(3\lambda -2)v^{2} +4}{\sqrt{1-b^{2} u_{R}^{2}}}\Biggr]\\
         & +bu_{S}\Biggl[\sqrt{1-b^{2} u_{S}^{2}}\left( 3\lambda v^{4} +8v^{2} -4\right) +\frac{2b^{2} (1-\lambda )u_{S}^{2}}{\left( 1-b^{2} u_{S}^{2}\right)^{3/2}} +\frac{4(3\lambda -2)v^{2} +4}{\sqrt{1-b^{2} u_{S}^{2}}}\Biggr]\Biggr\}\\
         & -Ma\cdot \frac{2\lambda }{b^{2} v}\left(\frac{\lambda -b^{2} u_{R}^{2}}{\sqrt{1-b^{2} u_{R}^{2}}} +\frac{\lambda -b^{2} u_{S}^{2}}{\sqrt{1-b^{2} u_{S}^{2}}}\right) \\
         & +a^{2} \cdot \frac{(\lambda -1)\lambda ^{2}}{2b^{2}}\left(\frac{b^{3} u_{R}^{3}}{\sqrt{1-b^{2} u_{R}^{2}}} +\frac{b^{3} u_{S}^{3}}{\sqrt{1-b^{2} u_{S}^{2}}}\right) +\mathcal{O}\left( M^{3} ,M^{2} a,Ma^{2} ,a^{3}\right).
        \end{aligned}
\end{equation}
Thus result is consistent with that in Ref.~\cite{li2020finite} where the calculation formula Eq.~\eqref{deltaBA1} is adopted. As is expected, the computation in Ref.~\cite{li2020finite} is much more complicated than ours.

\subsection{Infinity-unreachable asymptotically nonflat SAS spacetime (rotating solution in conformal Weyl gravity)}
Aiming at providing an alternative to general relativity to solve some issues such as dark energy, dark matter, and cosmological constants, researchers proposed the conformal Weyl gravity \cite{mannheim1989exact,kazanas1991general,riegert1984birkhoff}. Mannheim $et\ al.$ have found exact solutions in fourth-order conformal Weyl gravity \cite{mannheim1990conformal,mannheim1991solutions,mannheim1992conformal,mannheim1994newtonian}, including the rotating solution \cite{mannheim1991solutions} which can be written as the following expression in the Boyer-Lindquist coordinates \cite{varieschi2014kerr}
\begin{equation}
  \begin{aligned}
      \mathrm{d} s^{2} = & -\left[ 1+\frac{\mu r}{r^{2} +a^{2}\cos^{2} \theta } -\mathcal{K}\left( r^{2} -a^{2}\cos^{2} \theta \right)\right]\mathrm{d} t^{2}  \\
       & +\frac{r^{2} +a^{2}\cos^{2} \theta }{r^{2} +\mu r+a^{2} -\mathcal{K}r^{4}}\mathrm{d} r^{2} +\frac{r^{2} +a^{2}\cos^{2} \theta }{1-\mathcal{K}a^{2}\cos^{2} \theta \cot^{2} \theta }\mathrm{d} \theta ^{2} \\
       &+\left[\left( r^{2} +a^{2} -\frac{\mu ra^{2}\sin^{2} \theta }{r^{2} +a^{2}\cos^{2} \theta }\right)\sin^{2} \theta +\mathcal{K}a^{2}\frac{r^{4}\sin^{4} \theta -\left( r^{2} +a^{2}\right)^{2}\cos^{4} \theta }{r^{2} +a^{2}\cos^{2} \theta }\right]\mathrm{d} \phi ^{2}\\
       & +2\frac{\mu ra\sin^{2} \theta +\mathcal{K}a\left[ a^{2}\left( r^{2} +a^{2}\right)\cos^{4} \theta -r^{4}\sin^{2} \theta \right]}{r^{2} +a^{2}\cos^{2} \theta }\mathrm{d} t\mathrm{d} \phi ,
      \end{aligned}
      \label{conweylmetric}
\end{equation}
where $ \mu =-M(2-3M\gamma )$ and $ \mathcal{K} =k+ \gamma ^{2} (1-M \gamma )/(2-3 M \gamma )^{2}$. $\gamma$ and $k$ are two small parameters required by conformal gravity, $M$ is the mass of the source, and $a$ denotes the rotation parameter. We recast the metric component corresponding to $\mathrm{d}r^2$ as
\begin{equation}
    g_{rr} =\frac{1+\frac{a^{2}\cos^{2} \theta }{r^{2}}}{1+\frac{\mu }{r} +\frac{a^{2}}{r^{2}} -\mathcal{K} r^{2}} =\frac{1+\frac{a^{2}\cos^{2} \theta }{r^{2}}}{1-\frac{2M-3M^{2} \gamma }{r} +\frac{a^{2}}{r^{2}} -\left[ k+\frac{\gamma ^{2}}{4} +\mathcal{O}\left( M^{3} ,M^{2} \gamma ,M\gamma ^{2} ,\gamma ^{3}\right)\right] r^{2}} .
    \label{grr3}
\end{equation}
When certain conditions are met for the parameters, there exist two roots ($r$) such that the denominator of Eq.~\eqref{grr3} becomes zero \footnote{For example, when $a=0$, $\gamma=0$ and $k=\Lambda/3$, Eq.~\eqref{conweylmetric} becomes the Schwarzschild de-Sitter metric. For $0<9 \Lambda M^{2}<1$, there exist two positive roots $r_{+}$ and $r_{++}$ of $1-2M/r-\Lambda r^2/3 $ such that $0<2M<r_{+}<3M<r_{++}$. The root $r_{+}=(2 / \sqrt{\Lambda}) \cos (\epsilon / 3+4 \pi / 3)$, with $\cos \epsilon=-3 M\sqrt{\Lambda}$, describes the event horizon, and the root $r_{++}=(2 / \sqrt{\Lambda}) \cos (\epsilon / 3)$ localizes the cosmological event horizon \cite{podolsky1999structure}.}. The larger root corresponds to the cosmological event horizon, rendering the metric~\eqref{conweylmetric} infinity-unreachable and making the conventional GWOIA method invalid.

We calculate the deflection angle of particles moving in the equatorial plane with the generalized GW method. Firstly, substituting metric~\eqref{conweylmetric} into Eq.~\eqref{dudphi} leads to
\begin{equation}
  \begin{aligned}
    \left(\frac{\mathrm{d} u}{\mathrm{d} \phi }\right)^{2} =& \frac{1-b^{2} u^{2}}{b^{2}} +M\cdot 2u\left(\frac{1-v^{2}}{b^{2} v^{2}} +u^{2}\right) +k\cdot \left(\frac{1-v^{2}}{b^{2} u^{2} v^{2}} +1\right) \\
    & +\mathcal{O}\left( M^{2} ,Ma,M\gamma ,Mk,a^{2} ,a\gamma ,ak,\gamma ^{2} ,\gamma k,k^{2}\right),
  \end{aligned}
    \label{conweyldudphi2}
\end{equation}
with which we obtain
\begin{equation}
  \begin{aligned}
    u( \phi ) = & \frac{\sin \phi }{b} +M\cdot \frac{1+v^{2}\cos^{2} \phi }{b^{2} v^{2}} +k\cdot \frac{b\sin \phi }{2v^{2}}\left( 1-\frac{1-v^{2}}{\tan^{2} \phi }\right) \\
    & +\mathcal{O}\left( M^{2} ,Ma,M\gamma ,Mk,a^{2} ,a\gamma ,ak,\gamma ^{2} ,\gamma k,k^{2}\right),
  \end{aligned}
    \label{uofphiconweyl}
\end{equation}
and
\begin{equation}
    \phi(u) = \begin{cases}
    \Phi(u), & \text{if  } \left| \phi \right| <\frac{\pi}{2}, \\
    \pi - \Phi(u) , & \text{if  } \left| \phi \right| >\frac{\pi}{2},
        \end{cases}
        \label{phiofubumblebee}
\end{equation}
where
\begin{equation}
    \begin{aligned}
    \Phi(u) = & \arcsin( bu) +M\cdot \frac{b^{2} u^{2} v^{2} -v^{2} -1}{bv^{2}\sqrt{1-b^{2} u^{2}}} +k\cdot \frac{b\left( b^{2} u^{2} v^{2} -2b^{2} u^{2} -v^{2} +1\right)}{2uv^{2}\sqrt{1-b^{2} u^{2}}} \\
    & +\mathcal{O}\left( M^{2} ,Ma,M\gamma ,Mk,a^{2} ,a\gamma ,ak,\gamma ^{2} ,\gamma k,k^{2}\right).
\end{aligned}
\end{equation}
Secondly, substituting metric~\eqref{conweylmetric} into Eqs.~\eqref{alpharralphaphiphi} and \eqref{betaMalpha2} yields the corresponding $\alpha_{rr}$, $\alpha_{\phi\phi}$, and $\beta_\phi$. Thirdly, substituting $\alpha_{rr}$, $\alpha_{\phi\phi}$, $\beta_\phi$, and Eq.~\eqref{conweyldudphi2} into Eqs.~\eqref{Hr} and \eqref{Tr}, and using Eq.~\eqref{fphi}, we derive
\begin{equation}
  \begin{aligned}
    f(\phi )= & M\cdot \frac{\left( v^{2} +1\right)\sin \phi }{bv^{2}} -k\cdot \frac{b^{2}\left( 2-v^{2}\right)}{2v^{2}\sin^{2} \phi } +M^{2} \cdot \frac{\left( 2-v^{2}\right)^{2}\cos (2\phi )+3v^{4} +12v^{2}}{4b^{2} v^{4}}\\
     & -Ma\cdot \frac{2\sin \phi }{b^{2} v} +Mk\cdot \frac{b}{16v^{4}\sin^{3} \phi }\Big[\left( 2-v^{4} +v^{2}\right)\cos (4\phi )+3\left( 14-v^{4} -3v^{2}\right)\\
     & -12\left( 1-v^{2}\right)^{2}\cos (2\phi )\Big] +ak\cdot \frac{2b}{v\sin^{2} \phi } +\gamma ^{2} \cdot \frac{b^{2}\left( v^{2} -2\right)}{8v^{2}\sin^{2} \phi }\\
     & +k^{2} \cdot \frac{b^{4}\left[ 8-v^{4} -8v^{2} -2\left( 2-v^{2}\right)^{2}\cos (2\phi )\right]}{8v^{4}\sin^{4} \phi } +\mathcal{O}( M^{3} ,M^{2} a,M^{2} \gamma ,M^{2} k,Ma^{2} ,\\
     & Ma\gamma ,Mak,M\gamma ^{2} ,M\gamma k,Mk^{2} ,a^{3} ,a^{2} \gamma ,a^{2} k,a\gamma ^{2} ,a\gamma k,ak^{2} ,\gamma ^{3} ,\gamma ^{2} k,\gamma k^{2} ,k^{3}) ,
    \end{aligned}
\end{equation}
where $r_\gamma$ is expressed as the reciprocal of Eq.~\eqref{uofphiconweyl}. Finaly, substituting $F\left(\phi\right)$ (derived by integrating $f(\phi)$), $\phi_R= \pi-\Phi\left(u_R\right)$, and $\phi_S=\Phi\left(u_S\right)$ into Eq.~\eqref{deltaFF} yields the deflection angle of particles moving in the equatorial plane
\begin{equation}
  \begin{aligned}
    \delta = & M\cdot \frac{\left( v^{2} +1\right)\left(\sqrt{1-b^{2} u_{R}^{2}} +\sqrt{1-b^{2} u_{S}^{2}}\right)}{bv^{2}} +k\cdot \frac{b\left( v^{2} -2\right)}{2v^{2}}\left(\frac{\sqrt{1-b^{2} u_{R}^{2}}}{u_{R}} +\frac{\sqrt{1-b^{2} u_{S}^{2}}}{u_{S}}\right)\\
     & +M^{2} \cdot \left\{\frac{3\left( v^{2} +4\right) [\arccos (bu_{R} )+\arccos (bu_{S} )]}{4b^{2} v^{2}} +\frac{u_{R}\left[ 3v^{2}\left( v^{2} +4\right) -b^{2} u_{R}^{2}\left( 3v^{4} +8v^{2} -4\right)\right]}{4bv^{4}\sqrt{1-b^{2} u_{R}^{2}}}\right. \\
     & \left. +\frac{u_{S}\left[ 3v^{2}\left( v^{2} +4\right) -b^{2} u_{S}^{2}\left( 3v^{4} +8v^{2} -4\right)\right]}{4bv^{4}\sqrt{1-b^{2} u_{S}^{2}}}\right\} -Ma\cdot \frac{2\left(\sqrt{1-b^{2} u_{R}^{2}} +\sqrt{1-b^{2} u_{S}^{2}}\right)}{b^{2} v}\\
     & +Mk\cdot \frac{b}{2v^{4}}\left\{3\left( 1-v^{2}\right)^{2}\ln\left[\cot\frac{\arcsin (bu_{R} )}{2}\cot\frac{\arcsin (bu_{S} )}{2}\right] -\frac{1-2v^{2}}{\sqrt{1-b^{2} u_{R}^{2}}} -\frac{1-2v^{2}}{\sqrt{1-b^{2} u_{S}^{2}}}\right\}\\
     & +ak\cdot \frac{2}{v}\left(\frac{\sqrt{1-b^{2} u_{R}^{2}}}{u_{R}} +\frac{\sqrt{1-b^{2} u_{S}^{2}}}{u_{S}}\right) +\gamma ^{2} \cdot \frac{b\left( v^{2} -2\right)}{8v^{2}}\left(\frac{\sqrt{1-b^{2} u_{R}^{2}}}{u_{R}} +\frac{\sqrt{1-b^{2} u_{S}^{2}}}{u_{S}}\right)\\
     & +k^{2} \cdot \frac{b}{8v^{4}}\left[\frac{\left( v^{4} -8v^{2} +8\right)\left( 1+b^{2} u_{R}^{2} -2b^{4} u_{R}^{4}\right) +2v^{2} -4}{u_{R}^{3}\sqrt{1-b^{2} u_{R}^{2}}}\right. \\
     & \left. +\frac{\left( v^{4} -8v^{2} +8\right)\left( 1+b^{2} u_{S}^{2} -2b^{4} u_{S}^{4}\right) +2v^{2} -4}{u_{S}^{3}\sqrt{1-b^{2} u_{S}^{2}}}\right] +\mathcal{O} \left( M^{3} ,M^{2} a,M^{2} \gamma ,M^{2} k,Ma^{2} , \right.\\
     & \left. Ma\gamma ,Mak,M\gamma ^{2} ,M\gamma k,Mk^{2} ,a^{3} ,a^{2} \gamma ,a^{2} k,a\gamma ^{2} ,a\gamma k,ak^{2} ,\gamma ^{3} ,\gamma ^{2} k,\gamma k^{2} ,k^{3}\right) ,
    \end{aligned}
\end{equation}
which will reduce to the result in Ref.~\cite{li2020thefinitedistance}, i.e. Eq.~\eqref{daKerr} (the result of Kerr spacetime), when $\gamma=k=0$.

\section{Conclusion}
\label{conclusion}
The application and development of the existing GW method for SAS spacetimes are impeded by two issues. Firstly, for certain spacetimes with singular behavior, the infinite integral region is ill-defined. Secondly, the computation involved is cumbersome.
In this paper, we put forward a generalized GW method and the corresponding calculation formula to solve these issues. Specifically, with careful analysis to the integral of Gaussian curvature over the integral region and the integral of geodesic curvature along the auxiliary circular arc, we find that the radial coordinate of the auxiliary circular arc can be chosen arbitrarily. Hence the integral region without singular behavior can be constructed, and the ill-defined issue is solved. As for the complicated computation, based on the free choice of the auxiliary circular arc and the streamlining of the integral of geodesic curvature along the trajectory, we obtain a simplified formula, with which the deflection angle can be derived with few steps.
Finally, we compute the deflection angle of particles in Kerr spacetime and Kerr-like black hole in bumblebee gravity, the results are consistent with those from existing works, thereby convincingly validating the effectiveness and superiority of our method. Additionally, we present, for the first time, the deflection angle of particles for the rotating solution in conformal Weyl gravity.

In summary, the generalized GW method offers a comprehensive framework for describing the GW method for various scenarios while also substantially optimizing the related calculation. We believe that our work will greatly facilitate the application of the GW method in astrophysics.

\section*{Acknowledgments}
This work was supported in part by the National Key Research and Development Program of China Grant No. 2021YFC2203001 and in part by the NSFC (No. 11920101003, No. 12021003 and No. 12005016). Z. Cao was supported by “the Interdiscipline Research Funds of Beijing Normal University” and CAS Project for Young Scientists in Basic Research YSBR-006. This work was also supported in part by the National Natural Science Foundation of China (Grant No. 12205093).

\bibliography{refs}
\end{document}